\documentstyle[11pt,aaspp4,xray]{article}
\def\placefig#1{#1}
\begin{document}
\newcommand{\msun}{M_{\odot}}
\newcommand{\zsun}{Z_{\odot}}
\newcommand{\kms}{\, {\rm Km\, s}^{-1}}
\newcommand{\cm}{\, {\rm cm}}
\newcommand{\mpc}{\, {\rm Mpc}}
\newcommand{\seg}{\, {\rm s}}
\newcommand{\sr}{\, {\rm sr}}
\newcommand{\hz}{\, {\rm Hz}}
\newcommand{\hi}{H\thinspace I\ }
\newcommand{\hii}{H\thinspace II\ }
\newcommand{\heii}{He\thinspace II\ }
\newcommand{\heiii}{He\thinspace III\ }
\newcommand{\nhi}{N_{HI}}
\newcommand{\lya}{Ly$\alpha$\ }
\newcommand{\etal}{et al.\ }
\newcommand{\yr}{\, {\rm yr}}
\newcommand{\erg}{\, {\rm erg}}
\newcommand{\kev}{\, {\rm keV}}
\newcommand{\kelvin}{\, {\rm K}}
\newcommand{\ev}{\, {\rm eV}}
\newcommand{\eq}{eq.\ }
\newcommand{\jm}[1]{[JM: {\it #1} ] }
\newcommand{\ng}[1]{[NG: {\it #1} ] }
\newcommand{\uf}[1]{[UF: {\it #1} ] }

\load{\scriptsize}{\sc}
\def\ion#1#2{{\rm #1\,\sc #2}}
\def\OVIII{{\ion{O}{viii}}}
\def\OVII{{\ion{O}{vii}}}
\def\OVI{{\ion{O}{vi}}}
\def\NeVIII{{\ion{Ne}{viii}}}
\def\NeIX{{\ion{Ne}{ix}}}
\def\FeXVII{{\ion{Fe}{xvii}}}
\def\FeXX{{\ion{Fe}{xx}}}

\title{The X-Ray Forest:
A New Prediction of Hierarchical Structure Formation Models}
\author{Uffe Hellsten$^{1}$, Nickolay Y.\ Gnedin$^{2}$, and
Jordi Miralda-Escud\'e$^{3,4}$}
\affil{$^1$ Lick Observatory, University of California, Santa Cruz, CA 95064}
\affil{$^2$ Dept. of Astronomy, University of California, Berkeley, CA 94720}
\affil{$^3$ University of Pennsylvania, Dept. of Physics and Astronomy,
David Rittenhouse Lab.,
209 S. 33rd St., Philadelphia, PA 19104}
\affil{$^4$ Alfred P. Sloan Fellow}
\authoremail{ gnedin@tac-1.berkeley.edu, uffe@ucolick.org,
jordi@llull.physics.upenn.edu}

\begin{abstract}

  We use numerical simulations of structure formation in a Cold Dark
Matter model to predict the absorption lines in the soft X-rays produced
by heavy elements in the shock-heated intergalactic medium at low
redshift. The simulation incorporates a model for heavy element production
in galaxies and the subsequent dispersion of the metals to the
intergalactic medium. We analyze in particular absorption lines produced
by oxygen, and calculate the ionization stage taking into account the
observed X-ray background at the present time. We find that oxygen is
fully ionized by the X-ray background in low-density voids, and is mostly
in the form of $\OVII$ and $\OVIII$ in the sheets and filamentary regions.
Strong absorption lines of $\OVII$ and $\OVIII$ with equivalent widths
$W\sim 100 \kms$ are produced in filamentary regions of overdensities
$\sim 100$ and temperatures $\sim 10^6$ K, located in the outskirts of
groups and clusters of galaxies. The $\OVII$ line at $E=574$ eV is
generally the strongest one in these systems. Our model predicts that
any X-ray source (such as a quasar) should typically show about one
$\OVII$ absorption line with $W > 100 \kms$ in the interval from $z=0$
to $z=0.3$. These lines
could be detected with the upcoming generation of X-ray telescopes, and
their origin in intervening systems could be confirmed by the
association with groups of galaxies and X-ray emitting halos near the
line-of-sight at the same redshift. The hot intergalactic medium may
be one of the main reservoirs of baryons in the present universe, and
the heavy element X-ray absorption lines
offer a promising possibility of detecting this new component in the
near future.

\end{abstract}


\section{Introduction}

  Hierarchical theories of the formation of large-scale structure are
based on the hypothesis that the gravitational collapse of some type of
cold dark matter (i.e., any collisionless matter with a small enough
velocity dispersion to allow collapse down to scales much smaller than
the observed structure), triggered by initial density fluctuations,
is responsible for the formation of galaxies and for the later
assembly of groups, clusters, and superclusters of galaxies. This
generic scenario has been very successful in explaining a large variety
of observations, even though the origin of the initial density
fluctuations remains unknown, and simple models are
therefore assumed for the primordial power spectrum (generally based
on adiabatic, Gaussian, scale-invariant fluctuations).

  One of the predictions that have been made from these models, with
the use of numerical hydrodynamic simulations, is that despite the
wide range of scales over which the baryonic matter is able to
collapse (as dark matter halos successively merge from the onset of
non-linearity until the present time), a relatively large fraction of
baryons should still remain as ``intergalactic matter''. This
intergalactic medium should be forming a network of shock-heated gas
in the form of filamentary and sheet-like structures connected to
galaxy clusters and groups, as well as colder gas left out in voids,
as is revealed in numerical simulations (e.g., Ostriker \& Cen 1996).
A similar network of photoionized and shock-heated gas should also
be present at high redshift
(although at lower temperatures than at the present time, due to the
lower velocities of collapse at high redshift), and probably gives
rise to the hydrogen \lya forest (e.g., Cen \etal 1994,
Hernquist \etal 1996, Miralda-Escud\'e \etal 1996, Zhang \etal 1997,
1998).

  Here, a new prediction based on the same theory of hierarchical
clustering and the presence of intergalactic gas shall be studied.
It has already been shown observationally that heavy elements are
present in the high-redshift absorption systems (Songaila \& Cowie
1996 and references therein). It is very likely that many more
elements were spread to the intergalactic medium (hereafter, IGM)
at later times, either through galactic winds energized by supernova
explosions or active galactic nuclei (e.g., Dekel \& Silk 1986), or
simply by the process of gravitational merging, which can also lead
to some gas being ejected from halos back into the IGM (Gnedin \&
Ostriker 1997). As pointed out by Shapiro \& Bahcall (1980) and Aldcroft
\etal (1994), heavy elements in the intergalactic medium should cause
absorption lines (as well as continuum edges of absorption) on
background X-ray sources
In fact, for highly ionized, hot gas at $T\sim 10^6\kelvin$,
absorption lines from heavy elements are probably the only method of
detection, since hydrogen is highly ionized and its absorption lines
are very weak. The soft X-ray emission from low-density gas is also
very weak and would
generally be seen superposed with emission from other structures along
the line-of-sight, as well as the Galactic emission.
We shall denote these absorption lines by ``X-ray forest'',
in analogy to the \lya forest caused by hydrogen. The X-ray forest
is much more difficult to observe than the \lya forest,
due to the lower sensitivity and resolution in the X-ray band. The
possibility of observing this X-ray forest has also been discussed
by Aldcroft \etal (1994) and Canizares \& Fang (1998).

  Similar absorption lines may also be detectable in the
ultraviolet when the temperature is not very high; these are generally
caused by lithium-like ions like $\OVI$.
Mulchaey \etal (1996) proposed this as a method to detect halos of
hot has in poor groups of spiral galaxies, where the temperature of
the halo gas may be too low to have been detected in emission by ROSAT.
There is in fact evidence for a population of $\OVI$ absorbers among
the numerous absorption systems seen in quasar spectra that may arise
in hot, collisionally ionized gas (e.g., Burles \& Tytler 1996).

  In this paper we shall predict some properties of the X-ray forest
from a hydrodynamic simulation, and discuss the prospects for detecting
it in future X-ray missions.

\section{Simulations}

  We adopt a CDM+$\Lambda$ cosmological model in this investigation,
with parameters $\Omega_0=0.35$, $\Omega_{\Lambda}=0.65$, and $h=0.70$.
We adopt
the value of $\Omega_b=0.05$ for the cosmic baryon density, near the
upper limit implied by
the deuterium abundance (e.g., Burles \& Tytler 1998). We also
assume the Harrison-Zel'dovich $n=1$ spectrum of primordial fluctuations,
and we
normalize the model to {\it COBE} according to Bunn \& White (1997),
which gives us a value for $\sigma_8=0.97$, in agreement with
(and slightly on the high side of) cluster
abundances from Eke, Cole, \& Frenk (1996).
We use the linear Boltzmann code LINGER from the COSMICS package 
(Bertschinger 1995) to compute the linear transfer functions for our initial
conditions with sufficient accuracy.

The simulation has been performed with the SLH-P$^3$M cosmological
hydrodynamic code (Gnedin 1995; Gnedin \& Bertschinger 1996).
The physical modelling incorporated in the code is described in Gnedin
\& Ostriker (1997), and includes dynamics of the dark matter and cosmic
gas, evolution of the spatially averaged UV- and X-ray background radiation, 
star formation, stellar feedback, 
non-equilibrium ionization and thermal evolution of primeval 
plasma, molecular hydrogen chemistry, equilibrium metal cooling,
and self-shielding of the gas.

We have performed two runs with box sizes of 64 and 32 $h^{-1}{\rm\,Mpc}$,
and with $2\times64^3$ resolution elements and a dynamical range of 640,
in order to have a first estimate of the importance of the missing
large-scale power and the finite resolution of our simulations.

\section{Photoionization by the X-ray Background}

  Among the heavy elements that can produce absorption lines in the
X-rays, oxygen is the most abundant, so it is natural that it produces
the strongest lines. In this paper we shall focus our analysis on
oxygen lines. Previous work on
X-ray absorption by the IGM (Shapiro \& Bahcall 1980; Aldcroft \etal
1994) has shown that the elements producing the next strongest lines
are C, N, Ne and Fe; we shall comment only briefly on the absorption
lines of these species below. Oxygen in the
IGM should be photoionized by the observed X-ray background. The
spectrum of the X-ray background is well fitted by the following
formula (see Boldt 1987, Barcons \& Fabian 1992):
\begin{equation} J_x = J_0\, (\nu/\nu_X)^{-0.29}\,
\exp(-\nu/\nu_X) ~, \end{equation}
where $h\nu_X=40 \kev$, and the normalization is $J_0= 1.75 \times
10^{-26} \erg\cm^{-2}\seg^{-1}\sr^{-1}\hz^{-1}$.
A useful and simple example of the resulting ionization equilibrium
is the case of $\ion{O}{ix}$ and $\OVIII$, i.e., fully ionized
oxygen and the hydrogenic oxygen ion, in conditions of high
ionization where the abundance of $\OVII$ and any lower ionization species
is negligible. In this case, hydrogen and helium should be fully
ionized and charge transfer can be neglected. Since $\OVIII$ is a
hydrogenic ion, the photoionization cross section $\sigma_{\OVIII}$
and radiative recombination coefficient $\alpha_{\OVIII}$ are easily
related to the same quantities for hydrogen: $\sigma_{\OVIII}(\nu) =
\sigma_{HI}(\nu/64)/64$, and $\alpha_{\OVIII}(T) = 8 \alpha_{HI}(T/64)$.
The photoionization rate for $\OVIII$ is $\Gamma=4\pi \int d\nu \,
(J_{\nu}/h\nu)\sigma_{\OVIII} = 3.3\times 10^{-18} \seg^{-1}$.
The redshift evolution of the X-ray background is uncertain, depending
on the luminosity function of X-ray sources. Here we shall simply
assume $\Gamma\propto (1+z)^3$, which is reasonable if the present
sources make only a modest contribution to the background.
Defining the normalized electron density $\Delta_e$ as the ratio to
the mean electron density if all baryons were fully ionized,
$\Delta_e \equiv n_e/ \bar n_e$, with $\bar n_e = 2\times 10^{-7}\,
(1+z)^3\, (\Omega_b h^2/0.02) \cm^{-3}$, the recombination rate to
$\OVIII$ is $\alpha_{\OVIII}\, n_e \simeq 5\times 10^{-19}\, \Delta_e\,
(T/10^6 \kelvin)^{-0.7}\, (1+z)^3 \seg^{-1}$.

\def\capFA{
The fractional abundance of $\OVIII$ ({\it upper panel\/}) and $\OVII$ 
({\it lower panel\/}) as a function of
baryon number density for three values of temperature: 
$10^5\kelvin$ ({\it solid line\/}), 
$10^6\kelvin$ ({\it dashed line\/}), and 
$10^7\kelvin$ ({\it dotted line\/}). The vertical dashed line marks
the average cosmic density for the cosmological model under consideration
at $z=0.176$.
}
\placefig{
\begin{figure}
\insertfigure{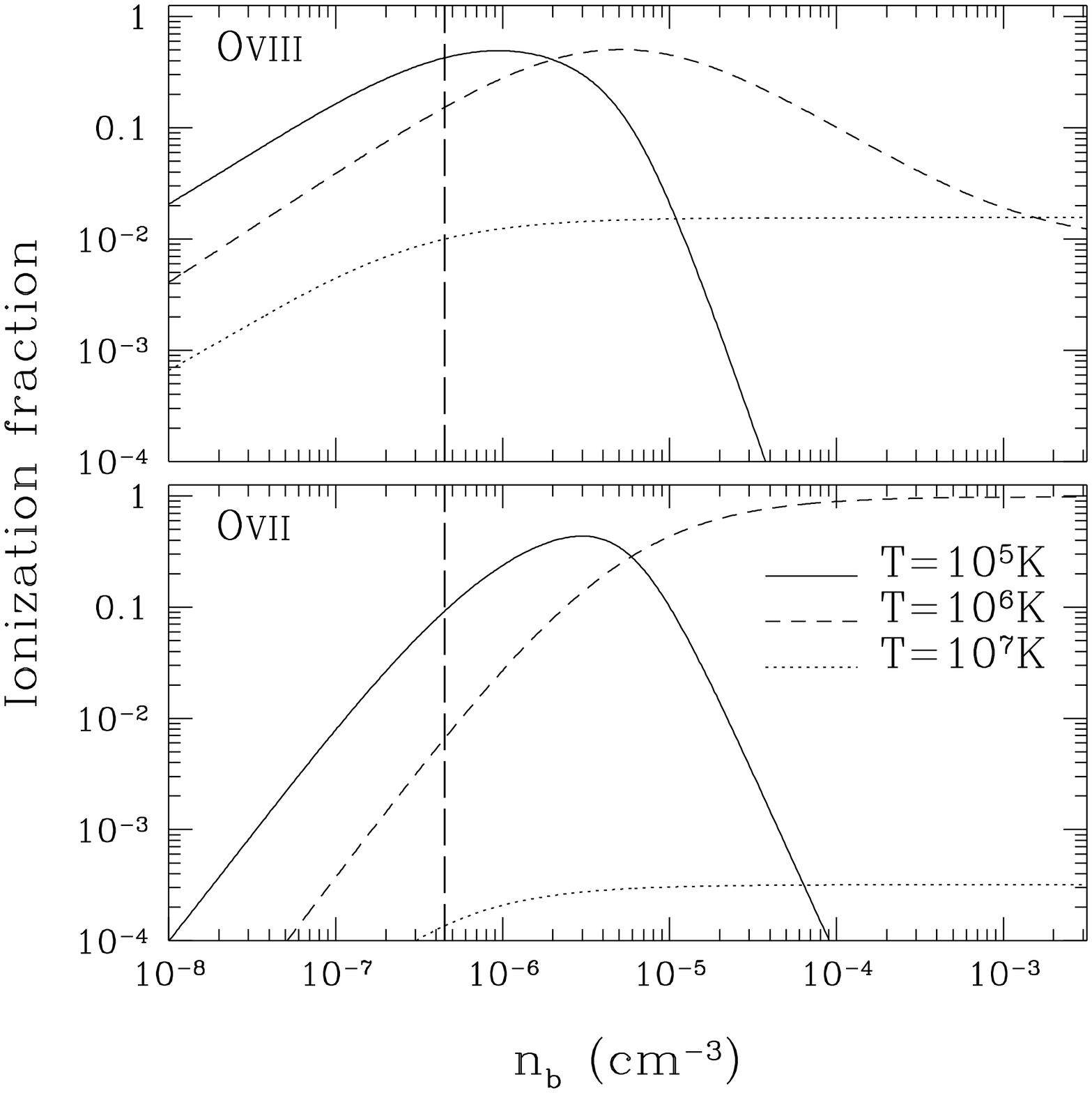}
\caption{\label{figFA}\capFA}
\end{figure}
}
We show in Figure 1 the fractional abundance of $\OVIII$ and $\OVII$ in
equilibrium, as a function of the baryon density, for the X-ray
background spectrum given above at $z=0.176$, and a gas temperature
$T=(10^5, 10^6, 10^7) \kelvin$ for the three curves plotted. These have
been calculated with the code CLOUDY 90 (Ferland 1996). The
mean baryonic density at $z=0.176$ is marked by a vertical dashed line.
Oxygen is mostly fully ionized in regions with density
$\Delta_e \lesssim 1$, due to photoionization from
the X-ray background. The fraction of $\OVIII$ at these low densities is
determined by the balance between photoionization and radiative
recombination, and is therefore proportional to the density, just like
for neutral hydrogen in the \lya forest. Collisional ionization at these
temperatures is negligible in comparison. At overdensities
$\Delta_e \simeq 3$, the $\OVIII$ fractional abundance reaches a maximum
and then starts to recombine to $\OVII$ at higher densities, so the $\OVIII$
fractional abundance is then inversely proportional to the density, if
oxygen is mostly in the form of $\OVII$.
At very high densities (or high temperatures), collisional ionization
dominates.

  We have assumed photoionization equilibrium to calculate the abundance
of $\OVIII$ throughout the simulation. Notice that this assumption is
actually not very good, since the photoionization time is of the order
of the Hubble time. A more accurate calculation would need to follow
the time-dependent evolution of the ionization fraction in every fluid
element as the simulation is run, but we have not implemented this
in the simulations presented here.

\section{Results}

\def\capSS{
A slice of the simulation volume from the 64 $h^{-1}\mpc$ run.
The slice is a sideway expansion (in a random
direction) of a random
line of sight, which is shown by white dotted line in each of five panels.
The slice dimension is $50h^{-1}$ comoving \mpc wide and $350h^{-1}$ 
comoving \mpc long.
The rows (from top to bottom) 
show $\OVII$ density, $\OVIII$ density,  metallicity,
temperature, and total gas density respectively.
}
\placefig{
\begin{figure}
\epsscale{1.0}
\insertfigure{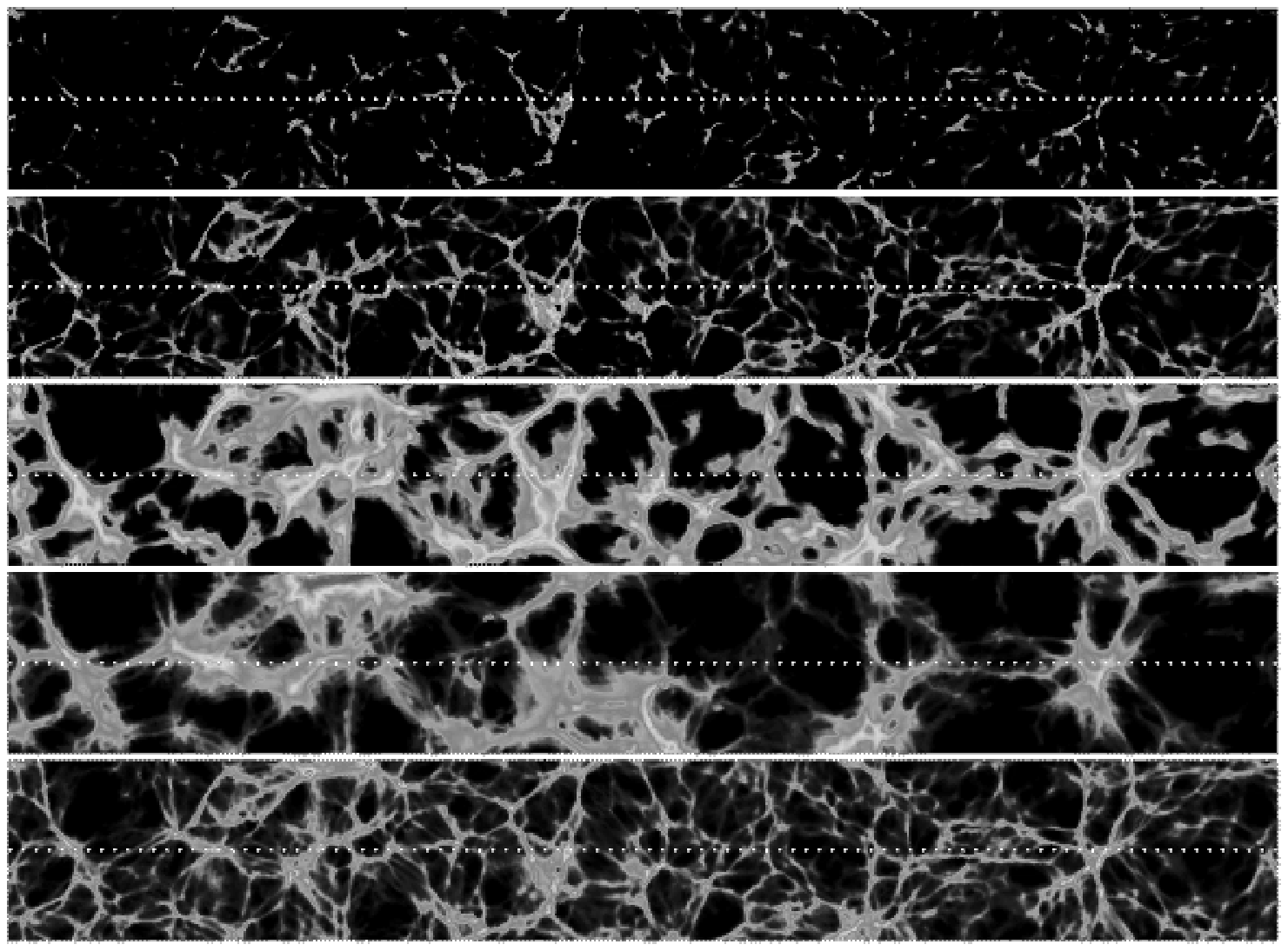}
\caption{\label{figSS}\capSS}
\end{figure}
}
A slice of the simulation in the $64 h^{-1}\mpc $ box at redshift
$z=0.176$ (of infinitesimal thickness) is shown in Figure 2 (Plate 1).
The five panels from the top to the bottom show $\OVII$ density, $\OVIII$
density, gas metallicity, temperature, and density.
The slice has been randomly selected: first, a random
line-of-sight (shown as a white dotted line in Fig. 2) along a random
direction through the box was chosen, and then a slice containing this
line-of-sight was selected at a random angle.
The width of the slice shown is $50h^{-1}{\rm\,Mpc}$,
and its length is $350h^{-1}{\rm\,Mpc}$.

The characteristic filamentary structure connecting collapsed clusters
is apparent. A fraction of about 50\% of the baryons in this
simulation are in regions with overdensities between 1 and 100,
corresponding to the filaments, with typical temperatures of $10^6
\kelvin$, and a fraction 6\% are at overdensities
between 100 and 1000, corresponding to the outer parts of cluster
halos, with typical temperatures $10^7 \kelvin$. 
The metallicity of
these regions is of order $0.1 \zsun$ in the simulation, caused by the
ejection of metals from halos during gravitational mergers (Gnedin \&
Ostriker 1997). In reality, possible mechanisms of enrichment of the IGM
are very uncertain, so the value of the metallicity should be considered
as a poorly known parameter, and of course the column densities of the
absorption lines we shall infer vary proportionally to the assumed
metallicity.
About 20\% of all baryons are converted to stars.

The distribution of $\OVIII$ is straightforward to interpret using
Figure 2: in general the $\OVIII$ density follows closely the filaments
of the density field, but the high-density clusters are not very
prominent due to the lower $\OVIII$ fraction at high densities. In
contrast, $\OVII$ is closely associated with the denser cluster halos.

\def\capLO{
A random line of sight through the slice from Fig.\ 2. The upper two
panels show $\OVII$ and $\OVIII$ fluxes respectively. The lower three
panels show gas metallicity, temperature and the total baryon
number density respectively.
}
\placefig{
\begin{figure}
\insertfigure{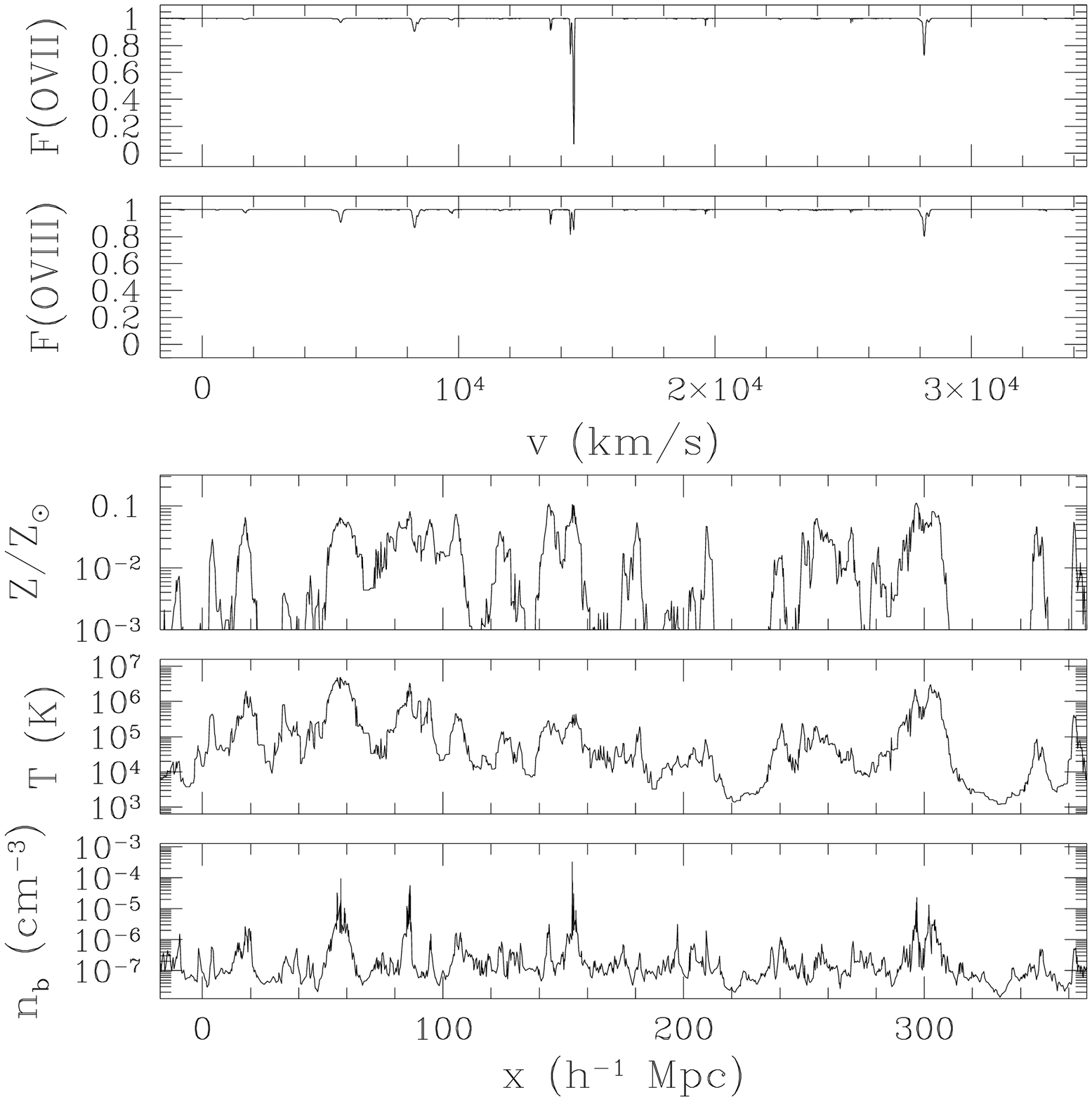}
\caption{\label{figLO}\capLO}
\end{figure}
}
In Figure 3, a line-of-sight through the slice of Figure 2 is
analyzed (this line-of-sight is the white dotted line in Figure 2). 
Shown in
Figure 3 are the gas density, temperature, and metallicity.
Since $\OVIII$ is hydrogenic, its strongest line is \lya with oscillator
strength $f=0.416$, at $\lambda =
19.0 \, {\rm \AA}$, or energy $653.6 \ev$, a good region for high
resolution X-ray spectroscopy. $\OVII$ has its strongest line (oscillator
strength $f=0.696$)
at $21.6 \, {\rm \AA}$ (energy $574.0 \ev$). The top two panels in
Figure 3 show the predicted spectrum produced by these two lines of
$\OVIII$ and $\OVII$. To compute the spectra, we use the formula
for the Gunn-Peterson optical depth for a uniform IGM
(Gunn \& Peterson 1965),
$\tau = \pi e^2/(m_ec)\, f\, H^{-1}(z) \lambda n(z)$, where $n(z)$ is
the density of the ion, and $H(z)$ the Hubble constant at redshift $z$,
and correct it for the effect of peculiar velocity gradient and
thermal broadening in the same way as in \lya forest simulations
(e.g., Cen \etal 1994).

  From the various absorption systems in Figure 3, we see the general
trend that the numerous weak systems have lines of similar strength in
$\OVIII$ and $\OVII$, while stronger absorption systems should be most
prominent in $\OVII$. This is also clearly seen in the two top panels of
Figure 2.
The strong absorbers correspond to higher density regions where $\OVII$
is more abundant than $\OVIII$.

  It needs to be emphasized that these X-ray absorption lines are
probing regions of a very different nature than the familiar X-ray
clusters detected from their X-ray emission. The temperatures of
the X-ray absorption systems are near $10^6$ K; the hotter gas
that is detected from X-ray emission ($T\gtrsim 10^7$ K) does not
produce much absorption because of the collisional ionization of
oxygen; in fact, even for a species that is not collisionally ionized
until higher temperatures, the cross sections of these hotter regions
are small. The X-ray absorbers arise from gas at the outskirts of the
most common virialized halos that are collapsing at the present time,
corresponding to poor groups and clusters of galaxies.

\def\capEW{
The distribution of equivalent widths of $\OVII$ ({\it solid lines\/}) and 
$\OVIII$ ({\it dashed lines\/}) absorption
lines in 
64 $h^{-1}\mpc$ box size ({\it bold lines\/}) and
32 $h^{-1}\mpc$ box size ({\it thin lines\/}) simulations.
}
\placefig{
\begin{figure}
\insertfigure{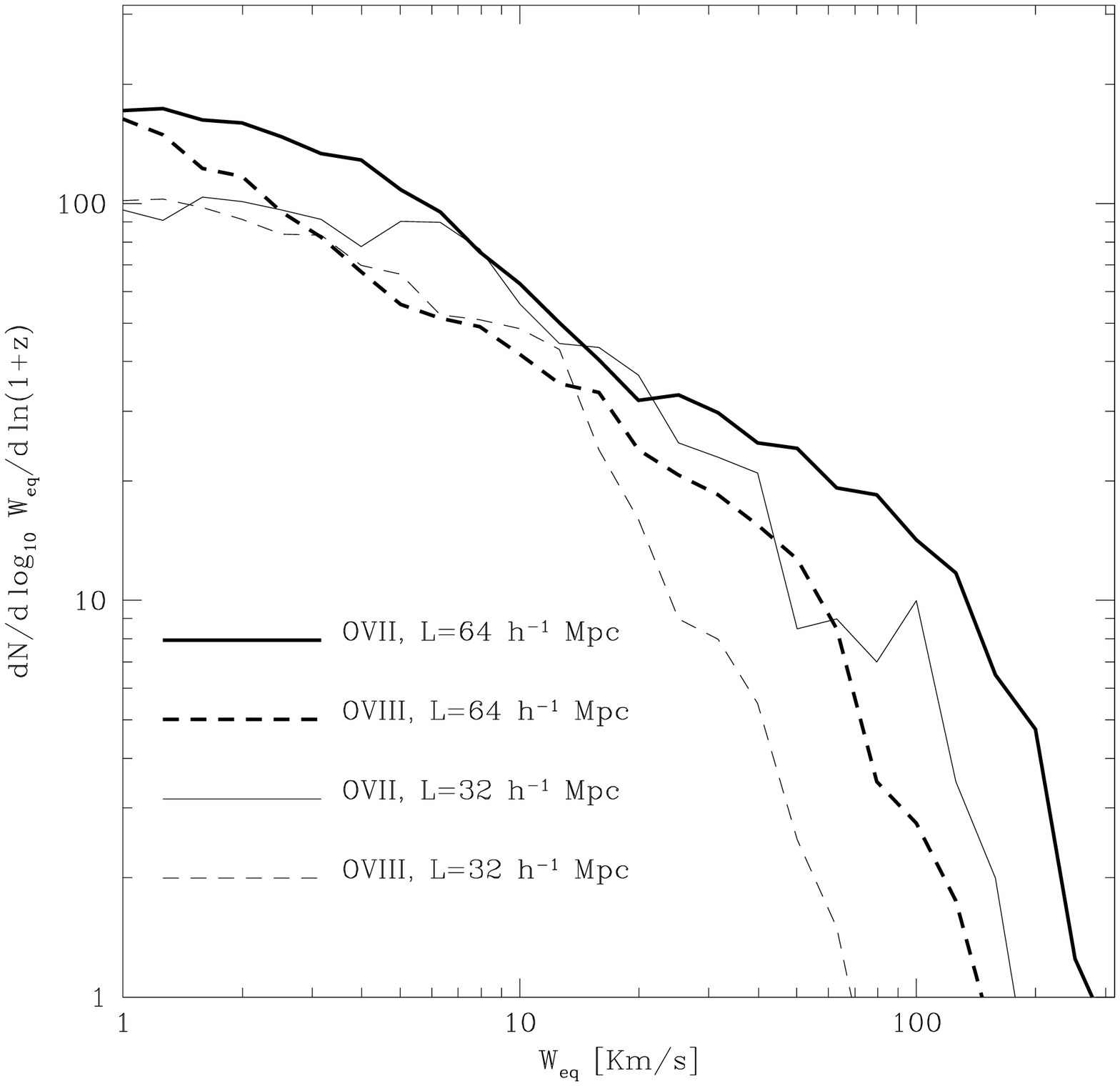}
\caption{\label{figEW}\capEW}
\end{figure}
}
  The distribution of equivalent widths of $\OVII$ and $\OVIII$ lines is
shown in Figure 4, for the two simulations with box size of $32$ and
$64 h^{-1} \mpc$. The main feature seen in the curves is that the
number of absorption systems falls sharply above an
equivalent width of $\sim 100 \kms$. The main reason for this
cutoff is line saturation, especially for the $\OVII$ lines. Even though
the velocity dispersion of the typical halos associated with the
absorption systems is larger than $100 \kms$, the fluid velocity
along a line-of-sight with high impact parameter has a dispersion
close to $100 \kms$, and thermal broadening is negligible due to the
high mass of the oxygen ion. Thus, the distribution of the $\OVII$ column
density is close to a power-law. On the other hand, the $\OVIII$ column
density distribution also shows a cutoff, although not as pronounced
as the cutoff for the equivalent width in Figure 4. The cutoff in
$\OVIII$ is therefore also due to an ionization effect, namely the fact
that high column density systems arise from regions of higher gas
density, where most of the oxygen has recombined to $\OVII$.

  The number of absorbers is larger in the $64 h^{-1} \mpc$ simulation
compared to the simulation on a smaller volume, for
$W \gtrsim 30 \kms$. The difference is mostly due to the missing
large-scale power in the small simulation, and the fact that the
volume in the small simulation is not large enough to contain
massive structures that can give rise to high equivalent width
systems at the resolution of the simulation.

  The results for the larger volume simulation in Figure 4 imply that
a random quasar at $z\sim 0.3$ should on average contain about one
$\OVII$ absorption line in its spectrum with $W > 100 \kms$, and about
four lines with $W > 30 \kms$. Given our comments above about the
effects of the finite volume of the simulation, we expect that the true
number of lines predicted by the cosmological models we have adopted
here should, if anything, be higher than the predictions from the
$64 h^{-1} \mpc$ box.
This equivalent width
is near the sensitivity limit of the upcoming X-ray missions AXAF and
Constellation X (Canizares \& Fang 1998).

  Many other absorption lines from other elements should be present
in the X-ray spectra of quasars (a list of permitted lines is given
by Verner, Verner, \& Ferland 1996). Every element should have the same
pair of lines we have considered here for oxygen (the \lya line of
the hydrogenic ion and the 1s$^2$-1s2p line of the helium-like ion).
There should also be permitted transitions of the K-shell electrons
in lithium-like ions (such as $\NeVIII$) that could be of interest, but
to our knowledge the oscillator strengths of these lines have not yet
been accurately computed (A. Pradhan 1998, priv. communication). 
The characteristics of these lines should be very similar to the
oxygen lines we have considered, but with reduced equivalent widths
due to the lower abundance. Lines at higher energies than the $\OVIII$ line
may be especially targeted by a mission like Constellation X,
where the sensitivity of the instruments improves at higher energy
(see http://constellation.gsfc.nasa.gov). The next strongest lines at
higher energies are $\NeIX$ (at $\lambda = 13.5 {\rm \AA}$),
$\FeXVII$ (at $\lambda=15.0 {\rm \AA}$) and
$\FeXX$ (at $\lambda = 12.8 {\rm \AA}$).
These lines are generally a factor $\sim$ 10 times weaker than the
$\OVII$ lines, for a fixed column density of gas and when the fraction
of the element in the observed ion is near its peak (the strength of
an unsaturated line is proportional to the oscillator strength,
times the abundance of the observed species, times the wavelength).
The lines of 
$\FeXVII$ and $\NeIX$ are present at similar temperatures as $\OVII$
and $\OVIII$, and should
generally trace similar regions, although they will be difficult to
detect given their weakness.
But $\FeXX$ is present up
to rather high temperatures (peaking near $10^7$ K), so some
particularly strong lines could occur on the rare lines of sight
intersecting massive clusters at small impact parameters, with
very large gas column densities.

\section{Conclusions}

  We have presented in this paper a new generic prediction of the
large-scale structure theories based on hierarchical gravitational
collapse of density fluctuations: X-ray quasars should show
absorption lines in their spectra from heavy elements in the intervening
intergalactic gas. A substantial fraction of the baryons are predicted
to be in the form of low-density intergalactic gas at the present time
(Ostriker \& Cen 1996; Cen \& Ostriker 1998; Miralda-Escud\'e \etal 1996;
Zhang \etal 1998).
Several mechanisms are known to enrich this gas with heavy elements;
given the observations of the metallicity in the centers of rich
clusters, it seems inevitable that this enrichment has also taken place
in the lower-density gas. In fact, the simulations presented here
{\it underpredict} the metal abundance observed in clusters; had we
increased the metal injection rate to fit the present cluster
metallicities, the column densities of the absorbers we predict would
be proportionally increased.

  Absorption lines in the soft X-rays, as well as far-UV lines produced
by lithium-like ions (Mulchaey \etal 1996), are the only observational
means we know of to detect low-density, hot intergalactic gas in the
present universe. At temperatures $\sim 10^6$ K, the neutral hydrogen
fraction is too low to produce significant \lya absorption, so we must
rely upon the highly ionized heavy elements.
The search
for intergalactic gas is important to complete an inventory of the
baryons in the present universe. The total baryon density observed
in galaxies and X-ray emitting gas in clusters comes close to the
baryon density predicted by nucleosynthesis (e.g.,
Persic \& Salucci 1997; Fukugita, Hogan, \& Peebles 1998), but given the
uncertainties it is possible for up to $\sim$ 80\% of the baryons to be
undetected, in the form of ionized intergalactic gas.
Results from cosmological simulations also suggest that the amount of
baryons in the \lya forest at $z\sim 3$ is higher than the sum of all
the baryons observed at present (Rauch \etal 1997; Weinberg \etal 1998;
Zhang \etal 1998).

  The detection and subsequent study of the X-ray forest would lead
to a large number of potential applications. The number of absorption
lines found will be measuring the product of the baryon density times
the metallicity in the IGM. The ratio of strengths of the $\OVII$ and $\OVIII$
lines will probe the distribution of the gas temperature and density,
and several other lines may be observed which could provide extensive
tests for the predictions of large-scale structure models. The X-ray
absorption lines could be correlated with structures seen in emission
near the line-of-sight, such as galaxies or X-ray emitting clusters
and groups. A new era in the study of the intergalactic medium could
be opened with the discovery of the X-ray forest by the new X-ray
observatories.

\acknowledgements

  We are grateful to Richard Mushotzky for useful discussions.
This work was supported in part by the 
UC Berkeley grant 1-443839-07427.
Simulations were performed on the NCSA Power Challenge Array under 
the grant AST-960015N.

\placefig{\end{document}}
 
\clearpage
 
\newcounter{figurecap}
\setcounter{figurecap}{0}
 
\begin{center}
\bf Figure Captions
\end{center}
 
\refstepcounter{figurecap}
Fig.\ \thefigurecap---\label{figFA}\capFA
 
\refstepcounter{figurecap}
Fig.\ \thefigurecap---\label{figSS}\capSS
 
\refstepcounter{figurecap}
Fig.\ \thefigurecap---\label{figLO}\capLO
 
\refstepcounter{figurecap}
Fig.\ \thefigurecap---\label{figEW}\capEW

\end{document}